\documentclass[runningheads]{svmult}

\usepackage{makeidx}   
\usepackage{graphicx}  
\usepackage{subeqnar}  
\usepackage{multicol}  
\usepackage{physprbb}  
\makeindex             

\begin{document}
\title*{GAIA -- Composition, Formation and Evolution of our Galaxy}

\toctitle{GAIA: Composition, Formation and Evolution of our Galaxy}
\titlerunning{GAIA: Astrophysics of the Galaxy}
%
\author{Gerry Gilmore}
\authorrunning{Gerry Gilmore}
\institute{Institute of Astronomy, Madingley Road, Cambridge CB3 0HA, UK}

\maketitle              

\begin{abstract}
GAIA will provide a multi-colour photometric and astrometric census of
some one billion compact sources, complete to 20th magnitude. In
addition, spectra for radial velocities will be obtained for about 30
million stars brighter than V=17. The high spatial resolution and
astrometric precision, 0.1arcsec and 10microarcsec to V=15, will not
only quantify the distribution of mass and the stellar populations in
the Galaxy, but make major advances in fundamental physics, cosmology
and solar system science. GAIA is an ESA mission, scheduled for launch
in mid-2010. A full description of the GAIA project and science case,
fairly crediting the hundreds of contributors, is available in the
project `Red Book' [ESA-SCI(2000)4]. A brief overview is provided by 
Perryman, de Boer, Gilmore, et al (2001). The www site is {\tt
http://www.rssd.esa.int/GAIA/}.

\end{abstract}

\section{Introduction}
Understanding the Galaxy in which we live is one of the great
intellectual challenges facing modern science. The Milky Way contains
a complex mix of stars, planets, interstellar gas and dust, radiation,
and the ubiquitous dark matter. These components are widely
distributed in age, reflecting their birth rate; in space, reflecting
their birth places and subsequent motions; on varied orbits,
determined by the gravitational force generated by their own and, more
importantly, the dark mass; and with chemical element abundances,
determined by the history of star formation, gas accretion, and mixing
in the ISM prior to their formation.  Astrophysics has now developed
the tools to measure these distributions in space, kinematics, and
chemical abundance, and to interpret the distribution functions to
map, and to quantify, the formation, structure, evolution, and future
of our entire Galaxy, given adequate data. 
This potential understanding is also of profound
significance for quantitative studies of the high-redshift Universe: a
well-studied nearby template underpins analysis of unresolved galaxies
at early times.

\subsection{Structure and Dynamics of the Galaxy}

The primary objective of the GAIA mission is the Galaxy: to observe
the physical characteristics, kinematics and distribution of stars
over a large fraction of its volume, with the goal of achieving
a full understanding of the Galaxy's dynamics and structure, and
consequently its formation and history. GAIA will make this goal
possible by providing, for the first time, a catalogue which will
sample a large and well-defined fraction of the stellar distribution
in phase space from which significant conclusions can be drawn for the
entire Galaxy. Hipparcos did this for one location in the Galaxy, the
Solar neighbourhood; GAIA will accomplish this for a large fraction of
the Galaxy.

\section{Space Astrometry}

The apparent motion of a star across the sky is dominated by proper
motion, which reflects the relative motion of the star and the Sun on
their Galactic orbits; parallax, which reflects the orbital motion of
the earth around our Sun, and provides a fundamental metric
calibration; and possible higher frequency terms which reflect
perturbations of the orbital motion of the target star from any
companions, such as planets. Since the amplitude of the earth's orbit
is known, the observed parallax determination of the angular parallax
may be converted into a metric distance. This however requires
extremely precise measurement, with precisions of a few
microarcseconds (pico-radians) required to probe beyond the immediate Solar
neighbourhood. Such precision is attainable in orbit, free from
atmospheric refractive problems, provided a suitably stable platform
is attainable. The ability of space astrometry to provide global
high-precision astrometry was proven by the ESA HIPPARCOS mission, one
of the greatest astrophysical advances of recent times.

The GAIA solution to this technical challenge is to mount two imaging
telescopes at a fixed angle on a single optical bench. Each telescope
has a wide field of view, in which the relative positions of all
sources are determinable. Since the telescopes are fixed on the same
bench, the relative positions of all sources in each telescope are
also determined relative to each other. As the satellite spins and
precesses, drift-scanning the sky across the large (180 CCDs) focal
plane, accurate relative positions of all objects are obtained. To fix
the zero point of the coordinates, all the observed unresolved quasars
are used to define a non-rotating (cosmological, Machian) reference
grid. The whole sky is observed every 70 days, providing, over a
5-year operational mission, an avearage of over 120 observations of
each source, sufficent data (4.6Mb/s for 5 years, or 0.1petabytes) to
model all of proper motion, parallax and multiplicity for all one
billion compact sources brighter than magnitude 20. That is one
percent of the stars in the Galaxy.

\subsection{ Relativistic astrometry} 

One of the more interesting aspects of precision measurement is that
the largest signal observed is not stellar motion, but general
relativistic distortion of the metric, even though GAIA will orbit at
L2, and not observe close to a line of sight near the Sun. 

\begin{table}[tbh]
\caption{Light deflection by masses in the Solar system. The monopole effect
dominates, and is summarized in the left columns for grazing incidence
and for typical values of the angular separation. Columns $\chi_{\rm
min}$ and $\chi_{\rm max}$ give results for the minimum and maximum
angles accessible to GAIA. J$_2$ is the quadrupole moment. The
magnitude of the quadrupole effect is given for grazing incidence, and
for an angle of 1$^\circ$. For GAIA this applies only to Jupiter and
Saturn, as it will be located at L2, with minimum Sun/Earth avoidance
angle of $\sim$50$^\circ$.}
\vskip 1pt
\label{tab:light-deflection}
\begin{center}
\leavevmode
\footnotesize
\begin{tabular}{|l|r|rrrr|r|r|r|}
\hline 
& \multicolumn{5}{|c|}{\null} &\multicolumn{3}{|c|}{\null} \\[-5pt]
Object        &\multicolumn{5}{|c|}{Monopole term} 
                  &\multicolumn{3}{|c|}{Quadrupole term} \\[5pt]
\cline{2-9} &&&&&&&& \\[-5pt]
&&&&&&&& \\[-5pt]
 & Grazing &$\chi_{\rm min}$ &$\chi=45^\circ$ &$\chi=90^\circ$
             &$\chi_{\rm max}$ &$J_2$ &Grazing 
                                                 &$\chi=1^\circ$ \\
 & $\mu$as  & $\mu$as & $\mu$as & $\mu$as & $\mu$as 
                                          &   & $\mu$as  & $\mu$as \\[5pt]
\hline  
&&&&&&&& \\[-5pt]
Sun     &1750000  &13000  &10000 &4100  &2100  &$\leq 10^{-7}$ & 0.3 & -- \\
Earth   &    500  &    3  &  2.5 & 1.1  &   0  &0.001 &  1 & --  \\ 
Jupiter &  16000  &16000  &  2.0 & 0.7  &   0  
                                       &0.015 &500 &$7\times 10^{-5}$ \\
Saturn  &   6000  & 6000  &  0.3 & 0.1  &   0  
                                       &0.016 &200 &$3\times 10^{-6}$ \\[5pt]
\hline
\end{tabular}
\end{center}
\end{table}

Table~\ref{tab:light-deflection} gives the magnitude of the deflection
for the Sun and the major planets, at different values of the angular
separation $\chi$, for the monopole term and the quadrupole term.
While $\chi$ is never smaller
than $50^\circ$ for the Sun (a constraint from GAIA's orbit), grazing
incidence is possible for the planets. With the astrometric accuracy
of a few $\mu$as, the magnitude of the expected effects is
considerable for the Sun, and also for observations near planets.

\section{Summary of the GAIA Science Capabilities}

Objectives: Galaxy origin and formation; physics of stars and their 
evolution; Galactic dynamics and distance scale; solar system census;
large-scale detection of all classes of astrophysical objects including 
brown dwarfs, white dwarfs, and planetary systems; fundamental physics

{\large \bf Measurement Capabilities:}
\begin{itemize}
\itemsep=0pt
\item {\bf catalogue:} $\sim1$~billion stars;
$0.34\times10^6$ to $V=10$~mag;
$26\times10^6$ to $V=15$~mag;
$250\times10^6$ to $V=18$~mag;
$1000\times10^6$ to $V=20$~mag; 
completeness to about 20~mag 

\item {\bf sky density:}
mean density $\sim25\,000$ stars deg$^{-2}$;
maximum density $\sim3\times10^6$ stars deg$^{-2}$ 

\item {\bf accuracies:} median parallax errors:
4~$\mu$as at 10~mag;
11~$\mu$as at 15~mag;
160~$\mu$as at 20~mag

\item {\bf distance accuracies:} from Galaxy models: 
21~million better than 1~per cent;
46~million better than 2~per cent; 
116~million better than 5~per cent; 
220~million better than 10~per cent\looseness=-2

\item {\bf tangential velocity accuracies:} from Galaxy models:
44~million better than  0.5~km~s$^{-1}$;
85~million better than  1~km~s$^{-1}$;
210~million better than  3~km~s$^{-1}$;
300~million better than  5~km~s$^{-1}$;
440~million better than 10~km~s$^{-1}$ 

\item {\bf radial velocity accuracies:} 1--10~km s$^{-1}$ to $V=16-17$~mag,
depending on spectral type

\item {\bf photometry:} to $V=20$~mag in 4~broad and 11~medium bands 

\end{itemize}

{\large \bf Scientific Goals:}
\begin{itemize}
\itemsep=0pt
\item {\bf the Galaxy:} 
origin and history of our Galaxy ---
tests of hierarchical structure formation theories --- 
star formation history --- 
chemical evolution --- 
inner bulge/bar dynamics --- 
disk/halo interactions --- 
dynamical evolution --- 
nature of the warp --- 
star cluster disruption --- 
dynamics of spiral structure --- 
distribution of dust ---
distribution of invisible mass ---
detection of tidally disrupted debris ---
Galaxy rotation curve ---
disk mass profile 

\item {\bf star formation and evolution:}
{\it in situ\/} luminosity function ---
dynamics of star forming regions --- 
luminosity function for pre-main sequence stars --- 
detection and categorization of rapid evolutionary phases --- 
complete and detailed local census down to single brown dwarfs --- 
identification/dating of oldest halo white dwarfs ---
age census ---
census of binaries and multiple stars

\item {\bf distance scale and reference frame}: 
parallax calibration of all distance scale indicators --- 
absolute luminosities of Cepheids --- 
distance to the Magellanic Clouds ---
definition of the local, kinematically non-rotating metric 

\item {\bf Local group and beyond:} 
rotational parallaxes for Local Group galaxies --- 
kinematical separation of stellar populations --- 
galaxy orbits and cosmological history --- 
zero proper motion quasar survey --- 
cosmological acceleration of Solar System ---
photometry of galaxies --- detection of supernovae 

\item {\bf Solar system:}
deep and uniform detection of minor planets ---
taxonomy and evolution ---
inner Trojans ---
Kuiper Belt Objects ---
disruption of Oort Cloud

\item {\bf extra-solar planetary systems:}
complete census of large planets to 200--500~pc --- 
orbital characteristics of several thousand systems

\item {\bf fundamental physics:} 
$\gamma$ to $\sim5\times10^{-7}$;
$\beta$ to $3\times10^{-4}-3\times10^{-5}$;
solar $J_2$ to $10^{-7}-10^{-8}$; 
$\dot G/G$ to $10^{-12}-10^{-13}$~yr$^{-1}$;
constraints on gravitational wave energy for $10^{-12}<f<4\times10^{-9}$~Hz;
constraints on $\Omega_{\rm M}$ and $\Omega_\Lambda$ from quasar microlensing

\item {\bf specific objects:} 
$10^6-10^7$ resolved galaxies; 
$10^5$~extragalactic supernovae;
$500\,000$ quasars;  
$10^5-10^6$ (new) solar system objects;
$\sim 50\,000$ brown dwarfs;
$30\,000$~extra-solar planets;
$200\,000$ disk white dwarfs;
$200$ microlensed events;
$10^7$ resolved binaries within 250~pc

\end{itemize}

\end{document}